\documentstyle[12pt,epsfig]{article}   

\def\lamb{{\mbox{\boldmath $\lambda$}}}
\def\EinLC{{\bf Ein}\kern-1em\raise0.4em\hbox{$^\circ$}\kern0.6em}
\def\RicLC{{\bf Ric}\kern-1em\raise0.4em\hbox{$^\circ$}\kern0.6em}
\def\delLC{\hbox{$\nabla$}\kern-0.6em\raise0.4em\hbox{$^\circ$}\kern0em}
\def\RFMLC{\hbox{$R$}\kern-0.3em\raise0.4em\hbox{$^\circ$}\kern0em}
\def\DDLC{\hbox{D}\kern-0.6em\raise0.4em\hbox{$^\circ$}\kern0em}
\def\RLC{\hbox{$\cal R$}\kern-0.6em\raise0.4em\hbox{$^\circ$}\kern0em}
\def\LL{{\cal L}}
\def\QQ{{\cal Q}}
\def\FF{{\cal F}}
\def\FFF{{\cal L}_1}
\def\E{{\cal E}}

\def\TTT{{\cal T}}
\def\hlambda{{\widehat{\lambda}}}

\def\TT{\ST_{[\bf{ric}]}{}  }
\def\td{\tilde}
\def\a{\alpha}
\def\b{\beta}
\def\dd{{\hbox{d}}}
\def\DD{{\hbox{D}}}
\def\g{{\gamma}} 
\def\gma{{\gamma}}
\def\r{\rho}
\def\th{\theta}
\def\ld{........}
\def\nml{{\cal N}}
\def\ax{{\cal A}}
\def\R#1#2{R^#1{}_#2}
\def\q#1#2{q^#1{}_#2}
\def\Q#1#2{Q^#1{}_#2}
\def\Om#1#2{\Omega^#1{}_#2}
\def\om#1#2{\omega^#1{}_#2}
\def\L#1#2{\Lambda^#1{}_#2}
\def\K#1#2{K^#1{}_#2}
\def\et#1#2{\eta^#1{}_#2}
\def\e#1{e^#1{}}
\def\sp#1{#1^{\prime}}
\def\l{\lambda}

\def\Tor{{\bf T}}
\def\dotTor{{\bf {\dot T}}}

\def\bfSS{{\bf S}}
\def\dotSS{{\bf \dot S}}

\def\nab#1{\nabla_{{}_#1}}
\def\i#1{i_{{}_{#1}}}
\def\ST{{\cal T}}
\def\pprime{^\prime}

\newcommand{\Ein}{{\rm \bf Ein}}
\newcommand{\bfg}{{\rm \bf g}}

\newcommand{\bfj}{\mbox{\boldmath $j$}}

\newcommand{\bfR}{\mbox{\rm \bf R}}

\newcommand{\calF}{{\cal F}}

\newcommand{\calS}{{\cal S}}
\newcommand{\calT}{{\cal T}}

\newcommand{\dcalM}{{\partial M}}
\newcommand{\del}{\mbox{$\nabla$}}
\newcommand{\efolding}{\mbox{$e$-folding}}

\newcommand{\itLamb}{\mbox{${\mit \Lambda}$}}
\newcommand{\Lie}{\pounds}
\def\modd{\hbox{\rm (mod \rmd)}}
\newcommand{\MM}{\hat{M}}
\newcommand{\NN}{\hat{N}}

\newcommand{\rinv}{\hat{\cal R}}
\newcommand{\rmd}{{\rm d}}
\newcommand{\rmi}{{\rm i}}
\newcommand{\rmK}{{\rm K}}

\newcommand{\scD}{\DDLC}
\newcommand{\Smin}{S_{\rm min}}
\def\SS{Y}
\def\und{\underbrace}
\def\var#1#2{ \und{\dot{#1} }_{#2}}

\newcommand{\wdg}{\mbox{$\,_\wedge\,$}}

\newcommand{\beqn}{\begin{equation}}	
\newcommand{\eeqn}[1]{\Eqno{#1}\vspace{2mm}\end{equation}}
\renewcommand{\(}	
	{\[\begin{array}{r@{\hspace{3pt}}c@{\hspace{3pt}}l}}
\renewcommand{\)}{\end{array}\vspace{2mm}\]}
\newcommand{\beqnarray} 	
	{\begin{equation}\begin{array}{r@{\hspace{3pt}}c@{\hspace{3pt}}l}}
\newcommand{\eeqnarray}[1]
	{\end{array}\Eqno{#1}\vspace{2mm}\end{equation}}
\newcommand{\dbox}[1]{\displaystyle #1}
\newcommand{\Eqno}[1]{\label{Eqn:#1}}
\newcommand{\Eqn}[1]{\mbox{(\ref{Eqn:#1})}}

\newcommand{\nno}{\mbox{\nonumber}}

\newcounter{Section}

\newcommand{\Section}[1]{
	\addtocounter{Section}{1}
	\setcounter{equation}{0}
	\vskip 50pt
	\begin{flushleft}
	{\large \bf \theSection.~#1}
	\end{flushleft}}

\begin{document}


\title{
DARK MATTER\\
GRAVITATIONAL INTERACTIONS
}

\author{Robin W Tucker\\
Charles Wang\\[0.8cm]
\it School of Physics and Chemistry, Lancaster University \\
\it   LA1 4YB, UK\\[0.5cm]
\tt r.tucker{\rm @}lancaster.ac.uk\\
\tt c.wang{\rm @}lancaster.ac.uk
}


\maketitle
\begin{abstract}



We argue that the conjectured dark mater in the Universe may be
endowed with a new kind of gravitational charge that couples to a
short range gravitational interaction mediated by a massive vector
field. A model is constructed that assimilates this concept into ideas
of current inflationary cosmology. The model is also consistent with
the observed behaviour of galactic rotation curves according to
Newtonian dynamics.  The essential idea is that stars composed of
ordinary (as opposed to dark matter) experience Newtonian forces due
to the presence of an all pervading background of massive
gravitationally charged cold dark matter. The novel gravitational
interactions are predicted to have a significant influence on
pre-inflationary cosmology.  The precise details depend on the nature
of a gravitational Proca interaction and the description of matter.  A
gravitational Proca field configuration that gives rise to attractive
forces between dark matter charges of like polarity exhibits
homogeneous isotropic eternal cosmologies that are free of
cosmological curvature singularities thus eliminating the horizon
problem associated with the standard big-bang scenario. Such solutions
do however admit dense hot pre-inflationary epochs each with a
characteristic scale factor that may be correlated with the dark
matter density in the current era of expansion.  The model is based on
a theory in which a modification of Einsteinian gravity at very short
distances can be expressed in terms of the gradient of the Einstein
metric and the torsion of a non-Riemannian connection on the bundle of
linear frames over spacetime. Indeed we demonstrate that the genesis
of the model resides in a remarkable simplification that occurs when
one analyses the variational equations associated with a broad class
of non-Riemannian actions.

\end{abstract}


\Section{Introduction}
The standard cosmological paradigm offers a fertile domain for the
fusion of ideas from astrophysics and particle physics. Certain
prejudices inherent in the classical Friedmann cosmologies are thought
to find a more natural resolution in terms of dynamical consequences
of phase transitions in the early Universe leading to a variety of
inflationary scenarios. Among the claims of success are the resolution
of {\it ad hoc} fine-tuning conditions and the puzzle of the observed
homogeneity and isotropy of matter. However a further problem is the
strong indication that the observed matter may be but a small fraction
of the gravitating matter in the Universe. In particular without a
significant contribution from directly unobserved matter it is hard to
reconcile Einstein's theory of gravitation (and its Newtonian limit)
with the dynamics of the Universe. Although knowledge of conditions in
the early Universe by their very nature require huge extrapolations
from the observed data, the conjecture that hidden matter also exists
seems necessary to explain the Newtonian dynamics of both individual
galaxies and galactic clusters.  For a star in a circular orbit of
radius $r$ with tangential speed $v$ within a spherical distribution
of galactic matter with mass $M(r)$, Newtonian dynamics predicts
$ v^2={G\, M(r)\over r}$ in terms of the Newtonian gravitational
coupling $G$.  Suppose $M_0\equiv M(r_0)$ is the constant mass of the
{\it visible} galaxy where $r_0$ denotes the radius of {\it visible}
matter.  If $r>r_0$ then in the absence of gravitating matter outside
the visible galaxy $ v^2(r)={{G\, M_0}\over r}$ i.e.
$v(r)\sim{\mbox{1}\over \sqrt{r}}$.  This however is not observed in
general.  For $r>r_0$, indications are (e.g from $H_\alpha$ emission)
that $ v^2(r)= k$ for some constant $k$, or $ M(r)\sim r $ requiring a
matter density distribution $\rho(r)$ of the form $ \rho(r)\sim{1\over
r^2}$. The material in
$M(r)$ for $r>r_0$ has been called {\it dark matter}. 

There have been many suggestions for the origin of these observations
 including a number that contemplate modifications of Newton's law of
 gravity and the effects of new interactions
 \cite{rosen,barnes,sherk,israelit_rosen,israelit,milgrom,mann,sciama}.
  This paper explores the idea that they result from the existence of
dark 
matter that is characterised by a novel gravitational interaction and that 
the carrier of this component of gravitation is a vector field that 
influences the pre-inflationary phase of the Universe
(c.f. \cite{ford}).
  As such it may modify the inflationary period needed to address the
so called ``flatness problem'' and may put the so called ``horizon
problem'' into a new perspective. 
Depending on the sign adopted for the coupling
of the Proca field in the fundamental action defing the theory this
 interaction may also be responsible for removing the classical 
curvature singularity associated with the traditional Friedmann
 cosmologies \cite{hawking_ellis}. 
 The coupling of the dark matter to the Proca component of gravitation
is defined by a new kind of conserved charge, the value of which
 can be correlated with the ``mass'' of the Proca quantum and the
 behaviour of the scale factor of the Universe. 

An important feature of this description is that it is rooted in
recent developments in non-Riemannian descriptions of
gravitation. 
Non-Riemannian geometries feature in a number of theoretical descriptions
of the interactions between fields and gravitation. Since the early pioneering
 work by Weyl, Cartan, Schroedinger, Trautman 
 and others such geometries have often
provided a succinct and elegant guide towards the search for unification of
the forces of nature
\cite{weylref} . In recent times interactions with supergravity have
been encoded into torsion fields induced by spinors and dilatonic
interactions from low energy effective string theories 
have been encoded into connections
that are not metric-compatible 
\cite{r1,r2,r3,r4}.
However theories in which the
non-Riemannian geometrical fields are dynamical in the absence of matter
are more elusive to interpret. 
We have   suggested  elsewhere \cite{newgravity}
that they may play an important role in certain astrophysical contexts.
Part of the difficulty in interpreting such fields is that there is little
experimental guidance available for the construction of a viable
theory that can compete effectively with general relativity in domains that
are currently accessible to observation. 
In such circumstances one must be
guided by the classical solutions admitted by theoretical models that admit
dynamical non-Riemannian structures
\cite{Hehl,Hehl2,Baekler,Ponomariev,McCrea,RWTCW,RWTCW_POLAND,Tey}.
This approach is being currently pursued by a number of groups
\cite{newgravity,RWTCW,RWTCW_POLAND,TresA,TresB,TresC,TresD,ob,neeman}. 
The work below continues in this vein by recognising that the essence
of the non-Riemannian approach is the existence of new gravitational
interactions. The nature of a particular interaction is explored in the
context of modern inflationary cosmologies and the dark matter problem.  
Although the dark matter model below is formulated in
terms of the standard Levi-Civita Riemannian connection the
introduction of the Proca component of gravitation is best appreciated
in terms of its genesis within non-Riemannian geometry.

\def\bfR#1#2{ {  R}_{{}_{#1,#2}}  } 

\def\dotbfR#1#2{ { \dot{R}}_{{}_{#1,#2}}  } 

\def\pprime{^\prime}
\def\frac#1#2{{#1\over #2}}
\def\ot{\otimes}
\def\CC{{\cal C}}
\def\RR{{\cal R}}

\Section{Non-Riemannian Geometry}
Einstein's theory of gravity has an elegant
formulation in terms of\break
 (pseudo-) Riemannian geometry.  The field equations
follow as a local extremum of an action integral under metric variations.
In the absence of matter the integrand of this action is simply the curvature scalar associated with
the curvature of the Levi-Civita connection times the (pseudo-) Riemannian
volume form of spacetime.  Such a connection $\nabla$ is torsion-free and
metric compatible.  Thus for all vector fields $X,Y$ on the spacetime
manifold, the tensors given by:
\begin{equation}
 \Tor(X,Y)=\nab{X}Y-\nab{Y}X-[X,Y] 
\end{equation}
\begin{equation}
\bfSS=\nabla \bfg 
\end{equation}
are zero,
where
 $ \bfg$ denotes the metric tensor, $\Tor$ the 2-1 torsion tensor 
and ${\bfSS}$ the gradient tensor of $\bfg$ with respect to $\nabla$.
Such a Levi-Civita connection provides a useful reference
connection since it depends entirely on the metric structure of the
manifold. This special connection will be denoted $\delLC$ in the following.


A general linear connection $\nabla$ 
on a manifold provides a covariant way to differentiate
tensor fields. It provides a type preserving derivation on the algebra 
of tensor fields that commutes with
contractions.  
Given an arbitrary local basis
of vector fields $\{X_a\}$ the most general
 linear  connection is specified locally
by a set of $n^2$ 1-forms  $\Lambda^a{}_b$ where $n$ is the dimension of the
manifold:
\begin{equation}
\nab{{X_a}} \,X_b=\Lambda^c{}_b (X_a)\, X_c . 
\end{equation}
Such a  connection can be fixed by specifying a (2, 0)
 symmetric metric tensor
$\bfg$,  a (2-antisymmetric, 1) tensor $\Tor$ and a (3, 0) tensor 
$\bfSS$, symmetric
in its last two arguments. If we require that $\Tor$ be the torsion of
$\nabla$ and $\bfSS$ be the gradient of $\bfg$ then it is straightforward to
determine the connection in terms of these tensors. Indeed since $\nabla$
is defined to commute with contractions and reduce to differentiation on
scalars it follows from the relation
\begin{equation}
X(\bfg(Y,Z))=\bfSS(X,Y,Z)+\bfg(\nab{X}Y,Z)+\bfg(Y,\nab{X}Z)
\end{equation} 
that 
\begin{eqnarray}
2\bfg(Z,\nab{X}Y)&=&X(\bfg(Y,Z))+Y(\bfg(Z,X))-Z(\bfg(X,Y))
\nonumber\\
&&{}-\bfg(X,[Y,Z])-\bfg(Y,[X,Z])-\bfg(Z,[Y,X])
\nonumber\\
&&{}-\bfg(X,\Tor(Y,Z))-\bfg(Y,\Tor(X,Z))-\bfg(Z,\Tor(Y,X))
\nonumber\\
&&{}-\bfSS(X,Y,Z)-\bfSS(Y,Z,X)+\bfSS(Z,X,Y)
\end{eqnarray}
where $X,Y,Z$ are any  vector fields.
The general curvature operator 
${\bf R}_{X,Y}$
defined in terms of $\nabla$
by
\begin{equation}
{\bf R}_{X,Y}Z=\nab{X}\nab{Y}Z-\nab{Y}\nab{X}Z-\nab{{[X,Y]}}Z 
\end{equation}
is also a type-preserving tensor derivation on the algebra of tensor fields.
The general (3, 1) curvature tensor ${\bf R}$ of $\nabla$ is defined by
\begin{equation}
{\bf R}(X,Y,Z,\beta)=\beta(\bfR{X}{Y} Z)
\end{equation}
where $\beta$ is an arbitrary 1-form. This tensor gives rise to a set of
local curvature 2-forms $R^a{}_b$: 
\begin{equation}
R^a{}_b(X,Y)=\frac{1}{2}\,{\bf R}(X,Y,X_b,e^a)
\end{equation}
where $\{e^c\}$ is any local basis of 1-forms dual to $\{X_c\}$.
In terms of the contraction operator $\rmi_X$ with respect to $X$
one has $\rmi_{X_b}\,e^a\equiv
\rmi_b\,e^a=e^a(X_b)=\delta^a{}_b$.
In terms of the connection forms
\begin{equation}
\R{a}{b}=\dd \Lambda^a{}_b+\Lambda^a{}_c\wedge \Lambda^c{}_b .
\end{equation}
In a similar manner the torsion tensor gives rise to a set of local torsion
2-forms $T^a$:
\begin{equation}
T^a(X,Y)\equiv \frac{1}{2}\, e^a(\Tor(X,Y)) 
\end{equation}
which can be expressed in terms of the connection forms as
\begin{equation}
T^a=\dd e^a+\Lambda^a{}_b\wedge e^b.
\end{equation}
Since the metric is symmetric the tensor $\bfSS$ can be used to define a set
of local non-metricity 1-forms $Q_{ab}$ symmetric in their indices: 
\begin{equation}
Q_{ab}(Z)=\bfSS(Z,X_a,X_b). 
\end{equation}

\Section{Non-Riemannian Actions}
\def\und{\underbrace}
\def\TT{\ST_{[{\bf ric}]}{}  }

\def\td{\tilde}
\def\a{\alpha\,}
\def\b{\beta\,}
\def\dd{{\hbox{d}}}
\def\DD{{\hbox{D}}}
\def\gma{{\gamma}}
\def\g{{\mbox{\boldmath $\gamma$}}}
\def\TT{{\mbox{\boldmath $\tau$}}}
\def\Cal{\cal}
\def\r{\rho}
\def\th{\theta}
\def\ld{........}
\def\nml{{\cal N}}
\def\ax{{\cal A}}
\def\R#1#2{R^#1{}_#2}
\def\q#1#2{q^#1{}_#2}
\def\Q#1#2{Q^#1{}_#2}
\def\Om#1#2{\Omega^#1{}_#2}
\def\om#1#2{\omega^#1{}_#2}
\def\L#1#2{\Lambda^#1{}_#2}
\def\K#1#2{K^#1{}_#2}
\def\et#1#2{\eta^#1{}_#2}
\def\e#1{e^#1{}}
\def\sp#1{#1^{\prime}}
\def\l{\lambda}
\def\Tor{{\bf T}}
\def\dotTor{{\bf {\dot T}}}
\def\bfSS{{\bf S}}
\def\dotSS{{\bf \dot S}}
\def\nab#1{\nabla_{{}_#1}}
\def\i#1{i_{{}_{#1}}}
\def\ST{{\cal T}}
\def\bfR#1#2{ {\bf R}_{{}_{#1,#2}}  } 
\def\dotbfR#1#2{ {\bf \dot{R}}_{{}_{#1,#2}}  } 
\def\bfg{{\bf g}}
\def\pprime{^\prime}
\def\frac#1#2{{#1\over #2}\,}
\def\ot{\otimes}
\def\CC{{\cal C}}
\def\k{\kappa\,}
\def\dott{\dot{\,}}
\def\dQ{\dd\,Q}
\def\var#1#2{ \und{(#1) }_{#2}\dott}
\def\Ein{{\bf Ein}}
\def\Ric{{\bf Ric}}
\def\Richat{{\bf \widehat{Ric}}}
\def\bfg{{\bf g}}
To exploit the geometrical notions above in a gravitational context
one must construct field equations that determine the torsion and
metric gradients.
These  are most naturally
derived from an action principle in which the metric, components of
the connection and matter fields
are the configuration variables. One requires that an action
be stationary with respect to suitable variations of such variables. 
If the action $4$-form
$\itLamb(\bfg,\nabla,\cdots)$ 
in spacetime  
contains the Einstein-Hilbert form
\beqn
\itLamb_{EH}(\bfg,\nabla)=\RR \star 1
\eeqn{EH}
where $\star$ denotes the Hodge map on forms
and $\underbrace{\dot{\itLamb_{EH}}}_{\bfg}$ denotes the variational derivative
of $\itLamb_{EH}(\bfg,\nabla)$ with respect to $\bfg $,
then
\beqn
\underbrace{\dot{\itLamb_{EH}}}_{\bfg}=
-h^{ab}\,\Ein(X_a,X_b)\,\star 1
\eeqn{xxxx}
where $h_{ab}\equiv\dot{\bfg}(X_a,X_b)$
and
\beqn
\Ein\equiv\Richat -\frac12 \bfg \RR
\eeqn{NRE}
is given in terms of the Ricci tensor
$\Ric (X,Y)={\bf R}(X_a,X,Y,e^a)$ by
\beqn
\Richat(X,Y)=\frac12(\Ric(X,Y)+\Ric(Y,X))
\eeqn{yyyy}
and $\RR=\Ric(X_a,X^a)$.
Further details of these variational calculations may be found in 
\cite{newgravity}.
Unlike the Einstein tensor associated with the Levi-Civita connection,
the Einstein tensor $\Ein$  defined in \Eqn{NRE}) is 
associated with the  non-Riemannian 
connection $\del$ and is not in general  divergenceless:
\beqn
(\del\cdot \Ein)(X_b) \equiv (\nabla_{X_a}\Ein)(X^a,X_b)
\neq 0.
\eeqn{zzz}
It is therefore instructive to decompose the non-Riemannian Einstein tensor
$\Ein$ into parts that depend on the Levi-Civita connection $\delLC$. For this
purpose  introduce the tensor $\lamb$  by
\beqn
\lamb(X,Y,\beta)=\beta(\nabla_{X}{Y})-\beta(\delLC_{X}{Y})
\eeqn{lambdef}
for arbitrary vector fields $X,Y$ and 1-form $\beta$.
In terms of the exterior covariant derivative \cite{benn_tucker} ${\DDLC}$
and  the Ricci tensor $\RicLC$ associated with the  Levi-Civita connection,
one may write:
\beqn
\Ric(X_a,X_b)=\RicLC(X_a,X_b)
+\rmi_{a}\,\rmi_{c}\,(
\DDLC\, \lambda{}^c{}_{b}+\lambda{}^c{}_{d} \wdg\lambda{}^d{}_{b})
\eeqn{xxx1}
where $
\lambda{}^a{}_{b}\equiv\lamb(-,X_b,e^a)
$ is a set of local 1-forms.
In terms of these forms
\beqn
T{}^a=\lambda{}^a{}_c\wdg e{}^c
\eeqn{T_lamb}
\beqn
Q{}_{ab}=-\lambda_{ab}-\lambda_{ba}.
\eeqn{Q_lamb}  
It follows from \Eqn{lambdef} that $\Ein$ differs from
the  Levi-Civita Einstein tensor $\EinLC$  by terms involving the
tensor $\lamb$ and its derivatives.
However for a large class of actions containing the torsion and metric
gradient fields one finds that these terms can be dramatically simplified.

To see this simplification most easily it is preferably to change
 variables from $\bfg , \nabla$ to $\bfg , \lamb$ in the total action
 $4$-form and write $
\itLamb(\bfg,\nabla,\cdots)=$\break $\LL(\bfg,\lamb,\cdots)
$. Since 
${\delLC}$ depends only on the metric
it follows from \Eqn{lambdef}
that 
$
\underbrace{\dot{\lamb}}_\nabla=\dot{\nabla}
$
and
$
\underbrace{\dot{\lamb}}_\bfg=
-\underbrace{\dot{\delLC}}_\bfg
$.
Hence the variational field equations are 
\vskip -10pt
\beqn
\underbrace{\dot{\itLamb}}_{\nabla}=
\underbrace{\dot{\LL}}_\lamb=0
\eeqn{delvar}
\beqn
\underbrace{\dot{\itLamb}}_{\bfg}=
\underbrace{\dot{\LL}}_\bfg
+\underbrace{\dot{\lamb}}_\bfg \underbrace{\dot{\LL}}_\lamb=0.
\eeqn{gvar}
From~\Eqn{delvar} one sees that
 the term $\underbrace{\dot{\LL}}_\lamb$ 
in \Eqn{gvar} does not contribute.
To evaluate these variations one first expresses the action in terms of $\bfg$
and the components of $\lamb$ and its derivatives. To this end it is
convenient to introduce the (traceless) 1-forms 
$
\hlambda^a{}_{b}\equiv\lambda{}^a{}_{b}
-{1\over 4}\,\delta{}^a{}_{b}\,\lambda{}^d{}_{d}
$, and the (traceless) 0-forms
$
\hlambda^a{}_{bc}\equiv\lambda{}^a{}_{bc}
-{1\over 4}\,\delta{}^a{}_{b}\,\lambda{}^d{}_{dc}
$
where $\lambda{}^a{}_{bc}\equiv \rmi_c\,\lambda{}^a{}_{b}$.
If we express the total action $4$-form $\LL(\bfg,\lamb, \cdots)$
as
\beqn
\LL(\bfg,\lamb, \cdots)=\LL_{EH}(\bfg,\lamb)+\FFF(\bfg,\lamb, \cdots)
\eeqn{action}
for some form $\FFF(\bfg,\lamb, \cdots)$
where
\beqn
\LL_{EH}=\RLC-\hlambda^a{}_c \wdg \hlambda^c{}_b \wdg \star  (e^b \wdg e_a)
-\dd(\hlambda^a{}_b \wdg \star (e^b \wdg e_a))
\eeqn{wrew}
in terms of the Levi-Civita scalar curvature $\RLC$\ \ then
\beqn
\underbrace{\dot{\LL_{EH}}}_\bfg
=-h^{ab}\,\EinLC(X_a,X_b)\,\star 1-h^{ab}\,\E_{ab}
\quad\quad\hbox{(mod d)}
\eeqn{tyurtyu}
where
\beqn
\E_{ab}=
{1\over2}\,\hlambda^q{}_d \wdg \hlambda^d{}_p \wdg
\{
g_{ab} \star  (e^p \wdg e_q) 
- \delta^p{}_a \star  (e_b \wdg e_q)
- \delta^p{}_b \star  (e_a \wdg e_q)
\}.
\eeqn{stuff}
Thus the Einstein field equation is
$$
\EinLC(X_a,X_b)\,\star 1+\E_{ab}+\TTT_{ab}=0
$$
where $
\underbrace{\dot{\FFF}}_{\bfg}=
-{h}^{ab}\,\TTT_{ab}$.
Next the variations with respect to $\lamb$ yield
\vskip -10pt
$$
\underbrace{\dot{\LL_{EH}}}_\lamb
=\dot{\hlambda}{}^a{}_b \wdg
\{\hlambda^c{}_a \wdg \star  (e^b \wdg e_c)
-\hlambda^b{}_c \wdg \star  (e^c \wdg e_a)\}
\quad\quad\hbox{(mod d)}
$$
\vskip -10pt
$$
\underbrace{\dot{\FFF}}_{\lamb}=
\dot{\lambda}^a{}_{b}\wdg \FF^{b}{}_a.
$$
By splitting off the trace part of the resulting field equation
one may write:
\beqn
\FF^a{}_a=0
\eeqn{cart1}
\beqn
\hlambda^c{}_a \wdg \star  (e^b \wdg e_c)
-\hlambda^b{}_c \wdg \star  (e^c \wdg e_a)+\widehat{\FF}^b{}_a=0
\eeqn{cart2}
where $
\widehat{\FF}^a{}_b\equiv {\FF}^a{}_b - {1\over 4}\delta^a{}_b\,\FF^c{}_c$.
To illustrate how the field equations 
\Eqn{cart1} and \Eqn{cart2} can greatly simplify
the terms in \Eqn{stuff}
 consider the eight parameter class of models in which
the torsion and metric gradient fields enter the action according to:
\beqnarray
\FFF&=&{4\,\kappa}\,R^a{}_a \wdg \star R^b{}_b
-2\,\ell\,\star 1+\alpha_1\,Q \wdg \star Q
\nno\\ 
&&
\hspace{-20pt}
+\alpha_2\,u \wdg \star  u
+\alpha_3\,v \wdg \star  v
+\alpha_4\,Q \wdg \star  u
+\alpha_5\,Q \wdg \star  v
+\alpha_6\,u \wdg \star  v
\eeqnarray{FF}
where $\kappa,\alpha_k$ are arbitrary coupling constants and $\ell$ is a
cosmological constant.
The 1-forms $u$ and $v$ may be expressed in terms of the torsion forms $T^a$
and non-metricity forms $Q_{ab}$ as follows:
$$
u\equiv \lambda^c{}_{ac}\,e^a=T-{1\over2}\,Q
$$$$
v\equiv\lambda_a{}^c{}_c\,e^a={1\over2}\,Q-{1\over2}\,\QQ-T
$$
where $
T\equiv \rmi_{a}T^a
$, $Q\equiv Q^a{}_a=-2\,\lambda^a{}_a$
and $
\QQ\equiv e^a\,i^{b} Q_{ab}
$.
Furthermore $R^a{}_a=-{1\over2}\dd Q$ is proportional to the Weyl field
2-form $\dd Q$.
Computing the variational derivatives above one finds that \Eqn{cart1}
yields:
\beqn
\dd \star   \dd Q
+{16\alpha_1-\alpha_4-\alpha_5\over16\kappa}\, \star   Q
+{8\alpha_4-2\alpha_2-\alpha_6\over16\kappa}\, \star   u
+{8\alpha_5-2\alpha_3-\alpha_6\over16\kappa}\, \star   v
=0
\eeqn{preproca}
while \Eqn{cart2} implies:
\beqn
u=\beta_1\,Q
\eeqn{u}
\beqn
v=\beta_2\,Q
\eeqn{v}
and
\beqnarray
\hlambda^a{}_{bc}&=&
\dbox{
-{4\beta_1+4\beta_2+1\over 24}\,\delta^a{}_b\,\rmi_c\,Q+ \\[10pt]
}
&&
\dbox{
{10\beta_2-2\beta_1+1\over 12}(\,g_{bc}\,i^a Q+\,\delta^a{}_c\,\rmi_b Q)
}
\eeqnarray{eqn}
where
$$
\beta_{{1}}={\frac {
6\,\alpha_{{3}}+3\,\alpha_{{4}}
-21\,\alpha_{{5}}-3\,\alpha_{{6}}
+54\,\alpha_{{3}}\alpha_{{4}}
-27\,\alpha_{{5}}\alpha_{{6}}-2
}
{
16-6\,\alpha_{{2}}-6\,\alpha_{{3}}+42\,\alpha_{{6}}
+27\,{\alpha_{{6}}}^{2}-108\,\alpha_{{2}}\alpha_{{3}}
}}
$$
$$
\beta_{{2}}=
{\frac 
{
6\,\alpha_{{2}}-21\,\alpha_{{4}}
+3\,\alpha_{{5}}-3\,\alpha_{{6}}
+54\,\alpha_{{2}}\alpha_{{5}}-27\,\alpha_{{4}}\alpha_{{6}}-2}
{
16-6\,\alpha_{{2}}-6\,\alpha_{{3}}+42\,\alpha_{{6}}
+27\,{\alpha_{{6}}}^{2}-108\,\alpha_{{2}}\alpha_{{3}}
}}.
$$
Substituting \Eqn{u} and  \Eqn{v} into
\Eqn{preproca}   one now obtains a Proca-type equation in the
form
\beqn
\dd \star   \dd Q
+\beta_3\, \star   Q
=0
\label{proca}
\eeqn{proca}
where
$$
\beta_{{3}}=
{16\alpha_1-\alpha_4-\alpha_5\over16\kappa}\, 
+{16\alpha_4-2\alpha_2-\alpha_6\over16\kappa}\,\beta_{{1}} 
+{16\alpha_5-2\alpha_3-\alpha_6\over16\kappa}\, \beta_{{2}}.
$$
From~the metric variation of $\FFF$ one finds
$$
\TTT_{ab}=
\ell\,g_{ab}\star  1+
\kappa\,\TT(X_a,X_b)\star  1
+\widehat{\TTT}_{ab}
$$
where
$$
\TT=\star  ^{-1}
\{
\beta_3\,(\rmi_a Q \wdg \star   \rmi_b Q -{1\over2}\,g_{ab}\, Q \wdg \star   Q)
+\rmi_a \dd Q \wdg \star   \rmi_b \dd Q -{1\over2}\,g_{ab}\, \dd Q \wdg \star   \dd Q\}\,
e^a \otimes e^b
$$
and
$$
\widehat{\TTT}_{ab}=
\zeta_1\,
g_{ab} Q \wdg \star   Q 
+
\zeta_2\,\rmi_a Q \wdg \star   \rmi_b Q
$$
in terms of
\beqn
\zeta_{{1}}=
{\frac {13\,\beta_{{1}}}{324}}
+{\frac {7\,\beta_{{2}}}{324}}
+{\frac {29\,\beta_{{1}}\beta_{{2}}}{162}}
+{\frac {23\,{\beta_{{1}}}^{2}}{324}}
-{\frac {{\beta_{{2}}}^{2}}{324}}
+{\frac {5}{1296}}
\eeqn{zeta1}
and               
\beqn
\zeta_{{2}}=
-{\frac {4\,\beta_{{1}}}{81}}
+{\frac {2\,\beta_{{2}}}{81}}
-{\frac {37\,{\beta_{{1}}}^{2}}{162}}
+{\frac {5\,\beta_{{1}}\beta_{{2}}}{81}}
+{\frac {11\,{\beta_{{2}}}^{2}}{162}}
-{\frac {1}{648}}.
\eeqn{zeta2}
Remarkably $\widehat{\TTT}_{ab}$ exactly cancels $\E_{ab}$ leaving the
Einstein equation in the form
\beqn
\EinLC+\ell\,\bfg+\kappa\,\TT=0.
\eeqn{ein}
Thus the  solutions to the field equations derived from
the action\break
$\LL_{EH}(\bfg,\lamb)+\FF$
where $\FF$ is given by
 \Eqn{FF}  may be generated from solutions to the 
 Levi-Civita Einstein-Proca
system \Eqn{proca}, \Eqn{ein} with arbitrary cosmological constant 
$\ell$. 
Once $Q$ is determined the full connection follows from
\beqn
\lambda^a{}_b=\widehat{\lambda}^a{}_b-{1\over 8}\delta^a_b \,Q
\eeqn{lamb_lambhat}
where $\widehat{\lambda}^a{}_b$ is given by \Eqn{eqn}.
The torsion and metric gradient then follow from \Eqn{T_lamb} and
\Eqn{Q_lamb}.

The above discussion has concentrated on the gravitational sector of
the theory. The inclusion of matter into $\FF$ is straightforward in
principle. For example if the matter is minimally coupled to the Weyl
form $Q$, the action $\FF$ will contain a term $Q\wdg {\bf j}$ where
the 3-form ${\bf j}$ is the Proca charged matter current.  Then
\Eqn{proca} will contain an extra source term proportional to ${\bf
j}$ while \Eqn{ein} will contain an addition stress tensor from the
variation of the matter action with respect to the metric tensor.

\Section{Divergence Conditions}

The previous section dealt with a general framework for deriving the
variational equations for matter interacting with non-Riemannian
gravitation. In the following we construct models in which the new
component of gravitation is described in terms of a Proca vector
field with the matter fields considered as thermodynamic fluids. 
Action principles for relativistic fluid models interacting with
gravitation require the imposition of sufficient constraints to
maintain consistency. Fortunately it is possible to write down the
coupled field equations for an Einstein-Proca-fluid system without
much effort by exploiting the Bianchi identities to ensure that the
fluids satisfy the required relativistic continuum field
equations. This procedure relies on recognising how the metric
variations of the components of the action are correlated with the
matter variations.

Suppose the underlying theory of gravitational and matter fields (with 
compact support)
is defined by an action $4$-form $\Lambda$ on 
spacetime $M$ as a sum of $4$-forms 
$\sum_{r}{\Lambda_r}$. Then in general there exists a set of
symmetric second degree tensors $\{\calT_s\}$, each member of which is
constructed from an element $\Lambda_s$ belonging to a subset
$\{\Lambda_s\}$ contained in $\{\Lambda_r\}$, satisfying $\delLC\cdot\calT_s=0$
in terms of the Levi-Civita connection $\delLC$. The subset 
$\{\Lambda_s\}$ is defined so that each member 
$\Lambda_s$ depends on 
a set of tensor fields $\{\Phi^s{}_k\}$ (and possibly their derivatives)
where the intersection of the sets $\{\Phi^s{}_k\}$
contains only the metric tensor on $M$.

The essence of this result is well known (although perhaps not widely
appreciated) and relies on the behaviour
of the action functional $\calS$ in response to variations of fields
induced by local diffeomorphisms of the manifold $M$. 
For any set of $4$-forms $\{\Lambda_r\}$
the action
\[
\calS=\int_M\Lambda=\sum_r\int_M{\Lambda_r}
\]
where 
$\Lambda$ is constructed from the metric tensor $\bfg$ and tensor
fields of arbitrary type $\{\Phi_1,\Phi_2,\cdots\}$ on $M$.  For any
local diffeomorphism on $M$ generated by a vector field $V$ \beqn
\int_{M}{\Lie_{V}{\Lambda_r}} =
\int_{M}{\rmd\,(\rmi_{V}\,{\Lambda_r})} =
\int_{\dcalM}{\rmi_{V}\,{\Lambda_r}} =0 \eeqn{id} since the Lie derivative
${\Lie_{V}}=i_V\rmd +\rmd \, i_V$ in terms of the contraction operator
$\rmi_V$, $\rmd \Lambda_r=0$ and all fields in $\Lambda_r$ are assumed to
have compact support.  To explicitly construct the set \{$\calT_s\}$
write the variations of $\Lambda_r$ with respect to $\Phi_k\in
\{\Phi^s{}_k\}$ as
\[
\underbrace{\dot{\Lambda_r}}_{\Phi_k}
=\dot{\Phi}_k\, \cdot \calF_k
\;\;\;\;\;\modd
\]
for some tensor $\calF_k$ contracted with $\dot{\Phi}_k$.
In an arbitrary local frame $\{X_a\}$ with dual coframe  $\{e^b\}$
such that $e^b(X_c)=\delta^b{}_c$, the  symmetric ``stress'' 
tensor, $\calT_r={{\calT}_r}{}_{ab}\,e^a\otimes\,e^b$,
associated with $\Lambda_r$ 
is obtained from  the metric variation:
\[
\underbrace{\dot{\Lambda_r}}_{\bfg}
=\dot{\bfg}(X_a,X_b)\,\calT_r{}^{ab}\,\star 1 
\;\;\;\;\;\modd .
\]
Since the Lie derivative is a derivation on tensors,
${\Lie_{V}{\Lambda_r}}$ may be expressed in terms of the field and
metric variations:
\( {\Lie_{V}{\Lambda_r}} &=&
(\Lie_V\,\bfg)(X_a,X_b)\,\calT_r{}^{ab}\,\star1 \\[10pt] && \;\;\; +
(\Lie_V\,\Phi_1)\,\cdot\calF_1 + (\Lie_V\,\Phi_2)\,\cdot\calF_2 +
\cdots \;\;\;\;\;\modd \) If the set $\{\Phi_1,\Phi_2,\ldots\}$ is
unique to the $4$-form $\Lambda_r$ then the equations $\{\calF_k=0\}$
constitute a set of variational field equations. If these are
satisfied then
\[
{\Lie_{V}{\Lambda_r}}
=
(\Lie_V\,\bfg)(X_a,X_b)\,\calT_r{}^{ab}\,\star1
\;\;\;\;\;\modd.
\]
Since
\[
(\Lie_V\,\bfg)(X_a,X_b)=
\rmi_a\delLC_{X_b}\,\widetilde{V}
+
\rmi_b\delLC_{X_a}\,\widetilde{V}
\]
where $\widetilde{V}=\bfg(V,-)$
and $\bfg$ is symmetric one may write
\(
\dbox{
{1\over2}\,
\int_{M}{\Lie_{V}{\Lambda_r}}
}
&=&
\dbox{
\int_{M}{
\calT_r{}^{ab}\,\rmi_a\delLC_{X_b}\,\widetilde{V}\,\star1 
}}. \\[10pt]
\)
For any  $\calT_r$ it is convenient to introduce the 
associated $3$-form  $J_{rV}$ by
\[
J_{rV}=\star\calT_r(V,-).
\] Now

\beqnarray
\rmd(J_{rV})&=&
\rmd(\calT_{rab}\,V^a\,\star e{}^b) \\[10pt]
&=&
\scD\,\calT_{rab}\wdg V^a\star e{}^b
+
\calT_{rab}\scD\,V^a\wdg\star e{}^b \\[10pt]
&=&
e^c\,\rmi_c\,\scD\,\calT_{ab}\wdg V^a\star e{}^b
+
\calT^{rab}\,e^c\,\rmi_c\,\scD\,V_a\wdg\star e{}_b \\[10pt]
&=&
(\delLC_{X_c}\,\calT_r)({X_a},{X_b})\,V^a\,g^{bc}\,\star1
+
\calT_r{}^{ab}\,\rmi_a\delLC_{X_c}\,\widetilde{V}\,\delta^c{}_b\,\star1 \\[10pt]
&=&
(\delLC\cdot\calT_r)(V)\,\star1
+
\calT_r{}^{ab}\,\rmi_a\delLC_{X_b}\,\widetilde{V}\,\star1 \\[10pt]
\eeqnarray{djv}
where  
$\scD$  
above denotes the exterior covariant derivative associated
with $\delLC$.
Thus for fields with compact support
\(
\dbox{
{1\over2}\,
\int_{M}{\Lie_{V}{\Lambda}}
}
&=&
\dbox{
\int_{M}{
\calT_r{}^{ab}\,\rmi_a\delLC_{X_b}\,\widetilde{V}\,\star1 
}} \\[10pt]
&=&
\dbox{
\int_{M}{
\rmd\,J_V
}
-
\int_{M}{
(\delLC\cdot\calT)(V)\,\star1
}} \\[10pt]
&=&
\dbox{
\int_{\dcalM}{
J_V
}
-
\int_{M}{
(\delLC\cdot\calT_r)(V)\,\star1 .
}}
\)
Since
$J_V=0$ on ${\dcalM}$ and $V$ is arbitrary it follows that
\[
\delLC\cdot\calT_r=0.
\]
This result follows from the definition of $\calT_r$ and the
imposition of the field equations and should not
be confused with a ``conservation law'' in an arbitrary spacetime. 
In general, ``conservation laws''
 owe their existence to further conditions.
For example if there exists a vector field  $V=K$ where $K$ satisfies
the Killing  condition 
(${\cal L}_K \bfg =0$) then both terms on the right hand side of
\Eqn{djv} are zero and $J_{rK}$ defines a genuine conserved current: 
\[
\rmd J_{rK}=0.
\]
Thus we have shown that every component $4$-form $\Lambda_r$ 
 that gives rise, by variation, to a set of field equations
$\{\calF_k=0\}$ for any of the dynamical 
tensor fields in the action, 
excluding the metric $\bfg$, 
also gives rise to a divergenceless second rank tensor 
$\calT_r$.



\Section{\bf The Einstein-Proca-Matter System}
Based on the reduction of the non-Riemannian action to a theory of
gravity in terms of the standard Levi-Civita torsion free, metric
compatible connection $\delLC$, we now construct a model of gravity
and matter that includes the Proca field in the gravitational
sector. As befits its origin in terms of purely geometrical concepts
the Proca field is regarded as a gravitational vector field that is
expected to modify the gravitational effects produced by the tensor
nature of Einsteinian gravity.  The model is based on \Eqn{ein} in
section 3 with $\ell=0$ and the right hand side replaced by the stress
tensors for fluid matter. We may normalise the Proca field $Q$ so that
the term $\kappa\,\TT$ becomes $\sigma \Sigma$ where $\Sigma$ is the
Proca stress tensor, $\sigma=\pm 1$ and the constant 
$\beta_3$ is replaced by the
square of the (real) Proca field mass $m_\alpha$.  In the following we
explore the consequences of both signs for $\sigma$.  With $\sigma=-1$
and $m_\alpha=0$ the model is analogous to the classical
Einstein-Maxwell-fluid system. With $\sigma=+1$ an interpretation of
$\sigma\,\Sigma$ as a stress tensor for matter would be ``unphysical''. In
the absence of other matter and in a weak field limit it might be
expected to lead to stability problems analogous to fermionic stresses
in the absence of quantisation. However
whatever criteria are needed to maintain the stability of the
Minkowski background in the presence of gravitational (metric) waves
might be similarly invoked in the presence of gravitational Proca fields.
We shall not immediately
discard the $\sigma=+1$ possibility since it has interesting consequences for
cosmology.
Henceforth we restore the physical
constants $G,c,\hbar$ and adopt MKS units when comparing with
observation.

In general we consider matter to be composed of  ordinary matter
defined to have zero
coupling to the Proca field and ``dark matter'' defined to have a non-zero
coupling. We call this coupling Proca charge and denote the basic unit
of Proca charge by $q$. For our discussion of
cosmology we model both types of matter by fluids with standard
stress tensors 
$\calT_0$ and $\calT_q$ respectively. Denoting the standard Levi-Civita
Einstein tensor by $\EinLC$ and the contribution of the Proca field
$\alpha$ to the Einstein equation by $\sigma\Sigma$, $\sigma=\pm1$
we have
\beqn
\EinLC+\sigma\,\Sigma={8\,\pi\,G\over c^4}\,(\calT_0+\calT_q)
\eeqn{eineq}
where
$$
\calT_0=(c^2\,\rho_0+P_0)\,V_0\otimes V_0+P_0\,\bfg
$$

$$
\calT_q=(c^2\,\rho_q+P_q)\,V_q\otimes V_q+P_q\,\bfg
$$

$$
\bfg(V_0,V_0)=-1
$$

$$
\bfg(V_q,V_q)=-1.
$$
The mass densities $\rho_f$, $f=0,q$ of ordinary and dark matter
 respectively are functions of the particle densities $n_f$ and the
  entropies $s_f$ per particle:
$$
\rho_0=\rho_0(n_0,s_0)
$$
$$
\rho_q=\rho_q(n_q,s_q).
$$
The pressure $P_f$ and temperature $T_f$ may be derived from Gibb's
relation 
\beqn c^2\,\rmd\rho_f= c^2\,\mu_f\,\rmd\,n_f+n_f\,T_f\,\rmd
s_f \eeqn{gibbs} 
where $\mu_f={\rho_f+P_f/c^2\over n_f}$ is the
associated chemical potential.  In terms of the notation introduced in
Section 3 we have rescaled the Weyl 1-form $Q$ to the 1-form $\alpha$
and $\TT$ to 
$$
\Sigma=
\left({c\,m_\alpha{}\over\hbar}\right)^2
\left(
\alpha\otimes\alpha-{1\over2}\,\alpha(\widetilde{\alpha})\,\bfg\right)
+\left(
\rmi_c\,F\otimes \rmi^c\,F-{1\over2}\,
\star^{-1}(F \wdg \star F)\,\bfg\right)
$$
where
$
F=\rmd\alpha
$
is the Proca field strength.
Equation \Eqn{eineq} for the metric must be supplemented by field equations for
the Proca field $\alpha$ and 
the fluid variables together with their equations of state.  In view of the
comments in Section 4 we
adopt as matter field equations
\beqn
\delLC\cdot\calT_0=0
\eeqn{div1}
and
\beqn
\delLC\cdot\left({8\,\pi\,G\over c^4}\,\calT_q-\sigma\,\Sigma\right)=0
\eeqn{div2}
which are certainly compatible with the Bianchi identity
$\delLC\cdot\Ein=0$ and give rise, as we shall demonstrate below, to the
expected
Lorentz forces on charged matter due to vector fields. 
Since the Proca field couples to a current $\bfj_q$ of
 Proca charged matter we have
\beqn
\rmd\star F+\left({c\,m_\alpha{}\over\hbar}\right)^2
\star\alpha+\sigma\,\bfj_q=0.
\eeqn{Procaeq}
The Proca charge current will be assumed to 
take the convective form
\beqn
\bfj_q=q\, n_q \star \widetilde{V_q}
\eeqn{jq}
with constant Proca charge, $q$ (of dimension length).

In a similar manner we assume that the Proca neutral particle 
current is given by
$$
\bfj_0= n_0 \star \widetilde{V_0}.
$$
We postulate  conservation of Proca charged particles
\beqn
\rmd(n_q\star\widetilde{V})=0
\;\;\; \Leftrightarrow  \;\;\;
\rmd \bfj_q=0  
\;\;\;
\mbox{(since $q$ is constant)}.
\eeqn{ccv}
If the neutral Proca matter is also conserved   (as befits behaviour
in the
late post inflationary epoch)
\beqn
\rmd \bfj_0=0
\eeqn{pcv}
and one may then interpret the matter field equations as equations of
motion for the fluid flows.

To explicitly develop these equations we must compute the 
 divergence of $\Sigma$. To facilitate this we introduce the tensors
\beqn
\Sigma_\alpha\equiv
\alpha\otimes\alpha-{1\over2}\,\alpha(\widetilde{\alpha})\,\bfg
\eeqn{siga}
and
\beqn
\Sigma_F\equiv
\rmi_c\,F\otimes \rmi^c\,F-{1\over2}\,
\star^{-1}(F \wdg \star F)\,\bfg
\eeqn{sigf}
so that $$\Sigma=
\left({c\,m_\alpha{}\over\hbar}\right)^2\Sigma_\alpha+\Sigma_F.$$
It is a standard result that
\beqn
\delLC\cdot\Sigma_F=\rmi_U\,F
\eeqn{dsigf}
where the vector field $U$ is defined by
$
\star\widetilde{U}=\rmd\star F.
$
To calculate the divergence of $\Sigma_\alpha$ 
consider
\beqnarray
\delLC\cdot(\alpha\otimes\alpha)
&=&
\delLC_{X_a}(\alpha\otimes\alpha)(X^a)
\\[10pt]
&\hspace{-40pt}=&\hspace{-20pt}
(\delLC\cdot\alpha)\,\alpha+\alpha^a\,\delLC_{X_a} \alpha
=
(\star^{-1}\rmd\star\alpha)\,\alpha+\alpha^a\,\delLC_{X_a} \alpha
\eeqnarray{delaa}
where $\alpha^a=\alpha(X^a)$ and the relation
$\delLC\cdot\beta=\rmi_{X^a}\delLC_{X_a}\beta=(-1)^{p+1}\,\star^{-1}\rmd\star\beta$
for any $p$-form $\beta$ has been used.
Furthermore, for any  $0$-form $f$
\beqn
\delLC\cdot(f\,\bfg)=\rmd\,f
\eeqn{latest}
since $\delLC$ is metric compatible. Hence
 \beqnarray
\delLC\cdot(\alpha(\widetilde{\alpha})\,\bfg) &=&
\rmd(\alpha(\widetilde{\alpha})) =
e^a\wdg\delLC_{X_a}(\alpha(\widetilde{\alpha})) \\[10pt] &=&
2\,(\delLC_{X_a} \alpha)(\widetilde{\alpha})\,e^a =
2\,\alpha^a\,\delLC_{X_a}
\alpha-2\,\rmi_{\widetilde{\alpha}}\,\rmd\,\alpha \eeqnarray{delaag}
where the vector field ${\widetilde{\alpha}}$ is the metric dual of
the form $\alpha$.  It follows from \Eqn{siga}, \Eqn{delaa} and
\Eqn{delaag} that
$$
\delLC\cdot\Sigma_\alpha=(\star^{-1}\rmd\star\alpha)\,\alpha
+\rmi_{\widetilde{\alpha}}\,F.
$$
Together with \Eqn{dsigf} the divergence of $\Sigma$ becomes
\beqn
\delLC\cdot\Sigma
=
\rmi_Z\,F+\left({c\,m_\alpha{}\over\hbar}\right)^2
(\star^{-1}\rmd\star\alpha)\,\alpha
\eeqn{lasttt}
where the vector field $Z$ is defined by
\beqn
\star\widetilde{Z}=\rmd\star F + \left({c\,m_\alpha{}\over\hbar}\right)^2
\star\alpha.
\eeqn{defZ}
Comparing \Eqn{defZ} and \Eqn{Procaeq} one has
\beqn
Z=\sigma\,q\,n_q \,V_q
\eeqn{lastttt}
while the exterior derivative of both sides of \Eqn{Procaeq}
yields
$$
\left({c\,m_\alpha{}\over\hbar}\right)^2\rmd\star\alpha
+\sigma\,\rmd \bfj_q=0.
$$
Therefore the divergence of $\Sigma$ becomes
$$
\delLC\cdot\Sigma=\sigma q\,n_q\,\rmi_{V_q}\,F
$$
using \Eqn{ccv}, \Eqn{lasttt} and \Eqn{lastttt}. 

For completeness we derive the divergence of the  generic fluid stress:
$$
\calT=(c^2\,\rho+P)\,V\otimes V+P\,\bfg
$$
where $\bfg(V,V)=-1$
and $\rho=\rho(n,s)$. It follows from \Eqn{latest} that
\beqn
\delLC\cdot\calT=(c^2\,\rho+P)\,\delLC_V \widetilde{V} + \rmd P + 
\widetilde{V}\,\star^{-1}\rmd ( (c^2\,\rho+P)\star\widetilde{V}).
\eeqn{lastt}
The third term on the right hand side of \Eqn{lastt} may be expanded as 
\beqnarray
\star^{-1}\rmd ( (c^2\,\rho+P)\star\widetilde{V})
&=&
\star^{-1}\{
\rmd (c^2\,\rho+P) \wdg \star \widetilde{V}
+(c^2\,\rho+P) \,\rmd\star\widetilde{V} \}
\\
&=&
\rmi_V \{\rmd (c^2\,\rho+P) + (c^2\,\rho+P)\star^{-1} \rmd \star \widetilde{V}\}
\eeqnarray{CCCCC}
by using the relation
$$
\beta \wdg \star \omega = (\rmi_{\widetilde{\beta}} \,\omega) \star 1
$$
for any 1-forms $\beta$ and  $\omega$.
Furthermore, the general conservation equation
$$
\rmd(n \star\widetilde{V})
=
\rmd n \wdg \star\widetilde{V} + n \rmd \star \widetilde{V}
=0
$$
implies that 
$$
\rmd \star \widetilde{V}=-{1\over n}\,\rmd n \wdg \star \widetilde{V}
=-{1\over n}\,(\rmi_V\,\rmd n)\star 1
$$
and so
$$
\star^{-1}\rmd\star\widetilde{V}=-{1\over n}\,(\rmi_V\,\rmd n).
$$
Thus \Eqn{CCCCC} becomes
$$
\star^{-1}\rmd ( (c^2\,\rho+P)\star\widetilde{V})
=
\rmi_V \left(\rmd (c^2\,\rho+P) - {(c^2\,\rho+P) \over n}\,\rmd n \right).
$$
Applying the thermodynamic relation \Eqn{gibbs}
to $\rho(n,s)$ gives
the divergence of $\calT$ finally as:
$$
\delLC\cdot\calT=(c^2\,\rho+P)\,\delLC_V\,\widetilde{V}+\Pi_V\,\rmd P 
+\rmi_V (n\, T \rmd s)\,\widetilde{V}
$$
where $
 \Pi_V\,\rmd P\equiv\rmd P +  (\rmi_V\rmd P)\widetilde{V}$
is the transverse part of $\rmd P$ with respect to $V$.
Thus the vanishing of the tangential components 
of the field equations \Eqn{div1} and \Eqn{div2}  
gives rise to the isentropic conditions:
\beqn
V_0(s_0)=0
\eeqn{vs0}
and
\beqn
V_q(s_q)=0
\eeqn{vsq}
while the vanishing of the  transverse components yields  the
appropriate relativistic Navier-Stoke's
type equations for neutral and charged fluids respectively.:
\beqn
(c^2\,\rho_0+P_0)\,\delLC_{V_0}\,\widetilde{V_0}=-{\Pi_{V_0}\,\rmd P}
\eeqn{fluid1}
\beqn
(c^2\,\rho_q+P_q)\delLC_{V_q}\,\widetilde{V_q}=-{\Pi_{V_q}\,\rmd P}
+{q\,n_q\,c^4\over 8\,\pi\,G}\,\rmi_{V_q}\,F.
\eeqn{fluid2}
In the case of negligible pressure ($P_f/c^2 \ll \rho_f,\; f=0,\,q$) 
the relation 
\Eqn{gibbs} yields $\rho_f\approx\mu_f\,n_f$ where the constant $\mu_f$ is
identified as the Newtonian
mass of the neutral and Proca charged  particle respectively. 
In the limit where Einstein gravity is negligible we may introduce coordinates
$\{t,\,x^1,\,x^2,\,x^3\}$
and write the metric
$\bfg\approx
-c^2\,\rmd t\otimes\rmd t
+\rmd x^1\otimes\rmd x^1
+\rmd x^2\otimes\rmd x^2
+\rmd x^3\otimes\rmd x^3$.
The Proca equation \Eqn{Procaeq} reduces to
\beqn
\left(
{\partial^2\over\partial x^2}+
{\partial^2\over\partial y^2}+
{\partial^2\over\partial z^2}
-\left({c\,m_\alpha{}\over\hbar}\right)^2
\right)A=\sigma\,q\,\hat{n_q}
\eeqn{wveq}
with the static ansatze $n_q=\hat{n_q}(x^k)$ and $\alpha=c\,A(x^k)\,\rmd t$
for $k=$ 1, 2, 3. The Green function associated with
\Eqn{wveq} is given by
\beqn
A=-{1\over4\,\pi\,r}\,e^{-{c\,m_\alpha\over\hbar}\,r}
\eeqn{Gfun}
where $r=\sqrt{(x^1)^2+(x^2)^2+(x^3)^2}$. It follows that $\sigma\,q\,A$
represents the
Proca field produced by a
single Proca charged source whose number density is 
 approximated by the Dirac delta distribution
$\hat{n_q}(x^k)=\delta^3(x^k)$.
If the world line of another particle 
has spatial velocity $\dot{x}^k(t)$ ($\dot{x^k}/c \ll 1$)
then
\beqn
\delLC_{V_q}\,{V_q}\approx
\ddot{x^1}\,{\partial\over\partial x^1}
+\ddot{x^2}\,{\partial\over\partial x^2}
+\ddot{x^3}\,{\partial\over\partial x^3}.
\eeqn{ddot}
If this particle also has Proca charge $q$ and mass 
$\mu_q$ then 
in the low energy and non-relativistic limit
\Eqn{fluid2} and  \Eqn{Gfun} imply that
$$
\mu_q\,\ddot{x^k}=
-{\sigma\,c^4\,q^2\over 8\,\pi\,G}\,{\partial\over\partial x^k}\,A.
$$
Thus the ``Yukawa potential energy'' $V(r)$ between these
two Proca charged particles with separation $r$ is
\beqn
V(r)=-{\sigma\,c^4\over32\,\pi^2\,G}\,
{q^2\,\over r}\,
e^{-{c\,m_\alpha\over\hbar}\,r}.
\eeqn{yukawa_pol}
Particles carrying the same sign  of Proca charge
attract or repel each other according as $\sigma=+1$ or $-1$ respectively,
with a force of  range  ${\hbar\over c\,m_\alpha}$.
We give arguments below why it is of interest to consider the possibility
that dark matter may interact with the
Proca field having the polarity $\sigma=+1$.


\Section{Implications from Galactic Dynamics}

The basic equations for our dark matter model are the Einstein
 equation \Eqn{eineq}, the matter field equations \Eqn{div1},
 \Eqn{div2}, the Proca equation \Eqn{Procaeq}, supplemented by the
 conservation equation \Eqn{ccv} for the Proca current \Eqn{jq} and
 the imposition of equations of state for thermodynamic variables
 satisfying the Gibb's relation \Eqn{gibbs}.  The dark matter is
 currently assumed to form an all pervading environment in which the
 less abundant ordinary matter moves. Since the ordinary matter is
 considered to have zero Proca charge its interaction with the dark
 matter is purely Newtonian in the non-relativistic low energy domain.
 We examine the extent to which this simple picture is sufficient to
 reconcile the galactic rotation curves with observation.  For
 simplicity we shall treat the stars in a galaxy to be test particles
 that experience Newtonian forces from a locally spherical
 distribution of low pressure ($P_q\ll\rho_q c^2$) dark matter gas 
 at some temperature $T$
 and pressure $P_q$.

Since the gas is charged one expects that its equation of state take
the {\em imperfect gas} form \cite{mcquarrie} 
\beqn
{P_q\over k_B\,T}=n_q+B_2\,n_q^2
+O(n_q^3)
\eeqn{virial}
where $k_B$ is Boltzmann's constant and
$B_2$ is the
second virial coefficient given in terms of
the Yukawa type potential \Eqn{yukawa_pol}.

Under approximately adiabatic conditions \Eqn{gibbs} yields
$\rho_q\approx\mu_q\,n_q$  with constant $\mu_q$.  The Newtonian
dynamics of the self gravitating dark matter gas sphere gives rise to
the Lane-Emden equation \beqn 4\,\pi\,r^2 \left({\partial
P_q\over\partial r}\right)=-4\,G\,\rho_q\,\MM \eeqn{leeq} where
$\MM=\MM(r)$ is the total mass within the sphere of radius $r$.  If
$q\not=0$ then in terms of the dimensionless quantities $b$, $\eta$,
$u$ and $\NN=\NN(u)$ defined by \beqn
\eta={4\,\sqrt{2}\,\pi\,G\,\mu_q\over c^2\,q} \eeqn{eta_mu} \beqn
u={32\,\pi^2\,G\,k_B\,T\over c^4\,q^2}\,r \eeqn{u_r} \beqn
\NN={1\over\mu_q}\MM(r(u)) \eeqn{N_M} and \beqn
b=32768\,{\pi^5\,G^3\,k_B{}^3\,T^3\,B_2\over c^{12}\,q^6} \eeqn{b_B2}
\Eqn{leeq} becomes \beqn u^3\left({\partial^2 \NN\over\partial
u^2}\right) +(\eta^2\,u\,\NN-2\,u^2)\left({\partial \NN\over\partial
u}\right) +b\left({\partial \NN\over\partial u}\right) \left(
{u\over2}\,{\partial^2 \NN\over\partial u^2} -{\partial
\NN\over\partial u} \right)=0. \eeqn{Heqn}

For $u \gg 1$ \Eqn{Heqn} has the asymptotic solution 
$\NN(u)\approx{2\,u\over\eta^2}$, corresponding to
\beqn
\MM(r)\approx{2\,k_B\,T\over\mu_q\,G}\,r
\eeqn{MMr}
irrespective of the sign of $\sigma$.
This gives rise to the asymptotic dark matter mass density distribution
\beqn
\rho_q(r)={k_B\,T\over2\,\pi\,G\,\mu_q\,r^2}
\eeqn{galac_den}
and the asymptotic dark matter pressure distribution
\beqn
P_q(r)={k_B^2\,T^2\over2\,\pi\,G\,\mu_q^2\,r^2}.
\eeqn{galac_press}
Thus  
$
{P_q/c^2\over\rho_q}={k_B\,T\over\mu_q\,c^2}
$
and so
the low pressure condition is satisfied if
$k_B\,T\ll \mu_q\,c^2$ (cold dark matter).
Within such a dark matter gas, the rotation speed for stars,
treated as test particles with zero Proca charge,
following circular orbits
of radius $r$ 
under Newtonian gravity
is approximately {\em constant} with value
\beqn
v_c=\sqrt{{G\,\MM(r)\over r}}\approx\sqrt{{2\,k_B\,T\over\mu_q}}.
\eeqn{vc}
This relation may be used to eliminate $k_B\,T\over\mu_q$ in 
\Eqn{galac_den} yielding the approximate dark matter density
\begin{eqnarray}
\rho_q(r)
&\!\!\approx\!\!&
\frac{{}v_c{}^2}{4\,\pi\,G\,r^2}.
\label{rhoqrvsq}
\end{eqnarray}
Taking a typical observational value \cite{peebles} $v_c\approx\,220\,{\rm
Km/sec}$ Equation (\ref{rhoqrvsq}) gives a value
$\rho_q(r)\approx6\times10^{-28}\;{\rm Kg/m^3}$ of the dark
matter density for $r=10\,{\rm kpc}$.

Although there is no fundamental reason why the dark matter should
have been in thermal equilibrium with the microwave background
radiation at any time we may bound our estimates by assuming that its
temperature now is no hotter than $T=2.7\, \rmK$.
For
$v_c\approx\,220\,{\rm Km/sec}$
 \Eqn{vc} 
implies that 
\beqn
\mu_q\approx 1.5\times10^{-23}\;{\rm Kg}
\eeqn{muvalue}
or
$\mu_q\,c^2\approx 10\,{\rm G\,eV}$.

\Section{Cosmological Models} 

We now turn attention to the implications for cosmology by
considering the class of  homogeneous and isotropic
 Robertson-Walker type metrics
\beqn
\bfg
=
-c^2\,\rmd t \otimes \rmd t 
+S^2\,\left(
\rmd \chi \otimes \rmd \chi
+f^2\,(
\rmd \theta \otimes \rmd \theta 
+\sin^2\theta\,\rmd\phi\otimes \rmd \phi)
\right)
\eeqn{rwg}
in local coordinates
$\{t,\, \chi,\,\theta,\,\phi\}$,
where $S=S(t)$ denotes the scale factor and the function $f=f(\chi)$
is specified by the curvature parameter $k$ according to
$$
f(\chi)=\sin \chi ,\;\;\;{\rm if}\;\; k=1, \;\;
{\rm(spatially\; closed\; universe)}
$$$$
f(\chi)=\chi,\;\;\;\;\;{\rm if}\;\; k=0, \;\;
{\rm(spatially\; flat\; universe)}
$$$$
f(\chi)=\sinh \chi,\;\;\;{\rm if}\;\; k=-1, \;\;
{\rm(spatially\; open\; universe)}
.$$
Both neutral and Proca charged  fluids are 
 supposed to have  the common velocity:
$$
V_0=V_q={1\over c}\,{\partial\over\partial t}
$$
to a first approximation.
Furthermore suppose that  each fluid maintains  the
same entropy so that $s_0$ and $s_q$ are constants. This is consistent
 with the (weaker) isentropic conditions \Eqn{vs0} and \Eqn{vsq}.  
The densities thus
reduce to $\rho_0=\rho_0(n_0)$ and $\rho_q=\rho_q(n_q)$ where $n_0=n_0(t)$
and $n_q=n_q(t)$.

We adopt as ansatze for the Proca 1-form  
\beqn
\alpha=A\,\rmd t
\eeqn{aat}
where
$A=A(t)$.
It follows that $F=\rmd\alpha=0$ and the Proca field equation 
\beqn
\rmd\star F+\left({c\,m_\alpha{}\over\hbar}\right)^2\star\alpha
+\sigma\,\bfj_q=0
\eeqn{Procafeq}
gives
\beqn
A={\sigma\,q\,n_q\,\hbar^2\over c^2\,m_\alpha{}^2}.
\eeqn{AA}
The 
Proca charge conservation equation
 \Eqn{ccv} yields
the general solution:
\beqn
n_q={N_q\over S^3}
\eeqn{nqs}
where
$N_q$
is a  dimensionless constant.

The Einstein equation \Eqn{eineq}
yields the differential equations
\beqn
\left({\partial S\over \partial t}\right)^2
-{8\,\pi\,G\over3}\,(\rho-\zeta_\alpha)\,S^2
+c^2\,k=0
\eeqn{eineq1}
and
\beqn
3\,{\partial^2 S\over \partial t^2}
+
4\,(\rho+3\,P-4\,\zeta_\alpha)=0
\eeqn{eineq2}
where the total mass density and 
total pressure are defined by
$$
\rho\equiv\rho_0+\rho_q
$$
$$
P\equiv P_0+P_q
$$
and
$$
\zeta_\alpha\equiv
{\sigma\,c^4\,m_\alpha{}^2\,A^2\over16\,\pi\,G\,\hbar^2}
.$$
Since $\rmd s_f=0$ for $f=0,\,q$, 
\Eqn{gibbs}
becomes 
\beqn
\rmd\rho_f=\mu_f\,\rmd\,n_f
\eeqn{gibb}
where
\beqn
\mu_f={\rho_f+P_f{}/c^2\over n_f}.
\eeqn{mu_rho}

It is instructive to write \Eqn{eineq1} and \Eqn{eineq2} in terms of  
the
Hubble parameter
\beqn
H(t)\equiv{1\over S(t)}\,{\partial S(t)\over\partial t}
\eeqn{hubble}
and the
deceleration parameter
\beqn
{q_d(t)}\equiv
-{1\over H(t)^2\,S(t)}\,{\partial^2 S(t)\over\partial t^2}.
\eeqn{decel}
Then  \Eqn{eineq1} and  \Eqn{eineq2} become
\beqn
\rho=\zeta_\alpha+{3\,H^2\over8\,\pi\,G}
+{3\,c^2\,k\over8\,\pi\,G\,S^2}
\eeqn{deneq}
\beqn
P={(2\,{q_d}-1)\,c^2\,H^2\over 8\,\pi\,G}
-{c^4\,k\over 8\,\pi\,G\,S^2}
+c^2\,\zeta_\alpha.
\eeqn{preseq}
Furthermore in terms of the critical density:
\beqn
\rho_{\rm c}(t)=\zeta_\alpha+{3\,H^2\over 8\,\pi\,G}
\eeqn{rhocrit}
we define
\beqn
\Omega\equiv
{\rho(t)\over\rho_{\rm c}(t)}
=
{\rho\over \zeta_\alpha-{3\,c^2\,k\over8\,\pi\,G\,S^2}}.
\eeqn{defOmega}
Using \Eqn{deneq} one may eliminate $\zeta_\alpha$ in \Eqn{defOmega}
to get
\beqn
\Delta\equiv
\Omega-1
=
{3\,c^2\,k\over 8\,\pi\,G\,\rho\,S^2-3\,c^2\,k}
\eeqn{defDelta}
This parameter  plays an important role in the discussion below.


\Section{Dynamical Modifications to Friedmann Cosmologies}

To study the the effect of  $\Sigma$ on cosmology we first consider 
equations of state for 
both Proca charged and neutral fluids to be approximated  by
 pressure-free dusts with conserved particle currents.
It then follows from $P_0=0$, $P_q=0$ and \Eqn{gibb} that
$$
\rho_0={\mu_0}\,n_0
$$
$$
\rho_q={\mu_q}\,n_q
$$
with the constant chemical potentials ${\mu_q}$, ${\mu_0}$
interpreted as the masses of the Proca charged and
neutral particles.
From~the particle conservation equation  \Eqn{pcv} 
 it follows that 
\beqn
n_0={N_0\over S^3}
\eeqn{n0s}
where $N_0$ is a dimensionless constant.
Hence from \Eqn{n0s} and \Eqn{nqs} one has 
\beqn
\rho_0={M_0\over\,S^3}
\eeqn{rho0}
\beqn
\rho_q={M_q\over\,S^3}
\eeqn{rhoq}
and \Eqn{AA} becomes
$$
A=
{\sigma\,q\,\hbar\,M_q\over \mu_q\,c^2\,m_\alpha^2\,S^3}
$$ 
where 
$M_0=N_0\,\mu_0$, $M_q=N_q\,\mu_q$.
Using \Eqn{gibb} it may be shown that \Eqn{eineq2} is satisfied provided $S$ is
not constant and the Equations \Eqn{n0s}, \Eqn{nqs} and \Eqn{eineq1} are
satisfied. Now  \Eqn{eineq1} becomes
\beqn
\left({\partial S\over \partial t}\right)^2
-{8\,\pi\,G\over3}\,\left({M\over S}-{\Theta\over S^4}\right)
+c^2\,k=0
\eeqn{keyeqn}
where 
\beqn
M\equiv M_0 + M_q.
\eeqn{MM}

\begin{figure}
\epsfig{file=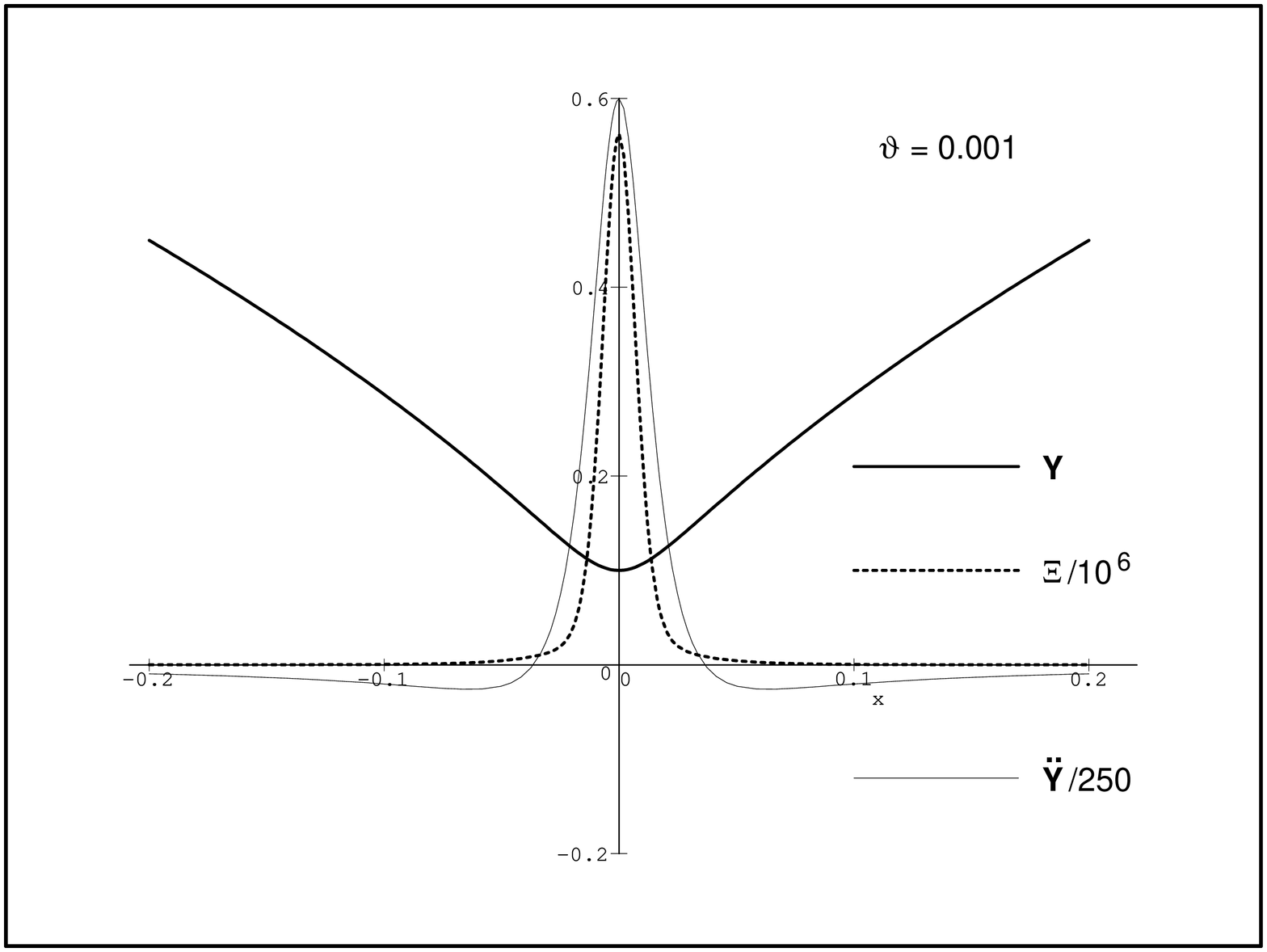,width=12cm,angle=-90}
\caption{The Early Universe for $\vartheta>0$}
\label{fig1}
The solution to \Eqn{keqn1} 
with a small positive value of $\vartheta$ has an asymptotic form
$\SS(x)\approx\vartheta^{1/3}+{3\over4}\vartheta^{-2/3}\,x^2$ around 
$x=0$. Such a $\SS(x)$ has a minimum at $x=0$, where the corresponding 
dimensionless
curvature invariant $\Xi(x)$ defined in \Eqn{defXi} 
has a maximal value. Note that
$\SS(x)$ has an acceleration phase ($\ddot{\SS}(x)>0$) within the region 
$-2\,(\vartheta/3)^{1/2}<x<2\,(\vartheta/3)^{1/2}$.
\end{figure}

\begin{figure}
\epsfig{file=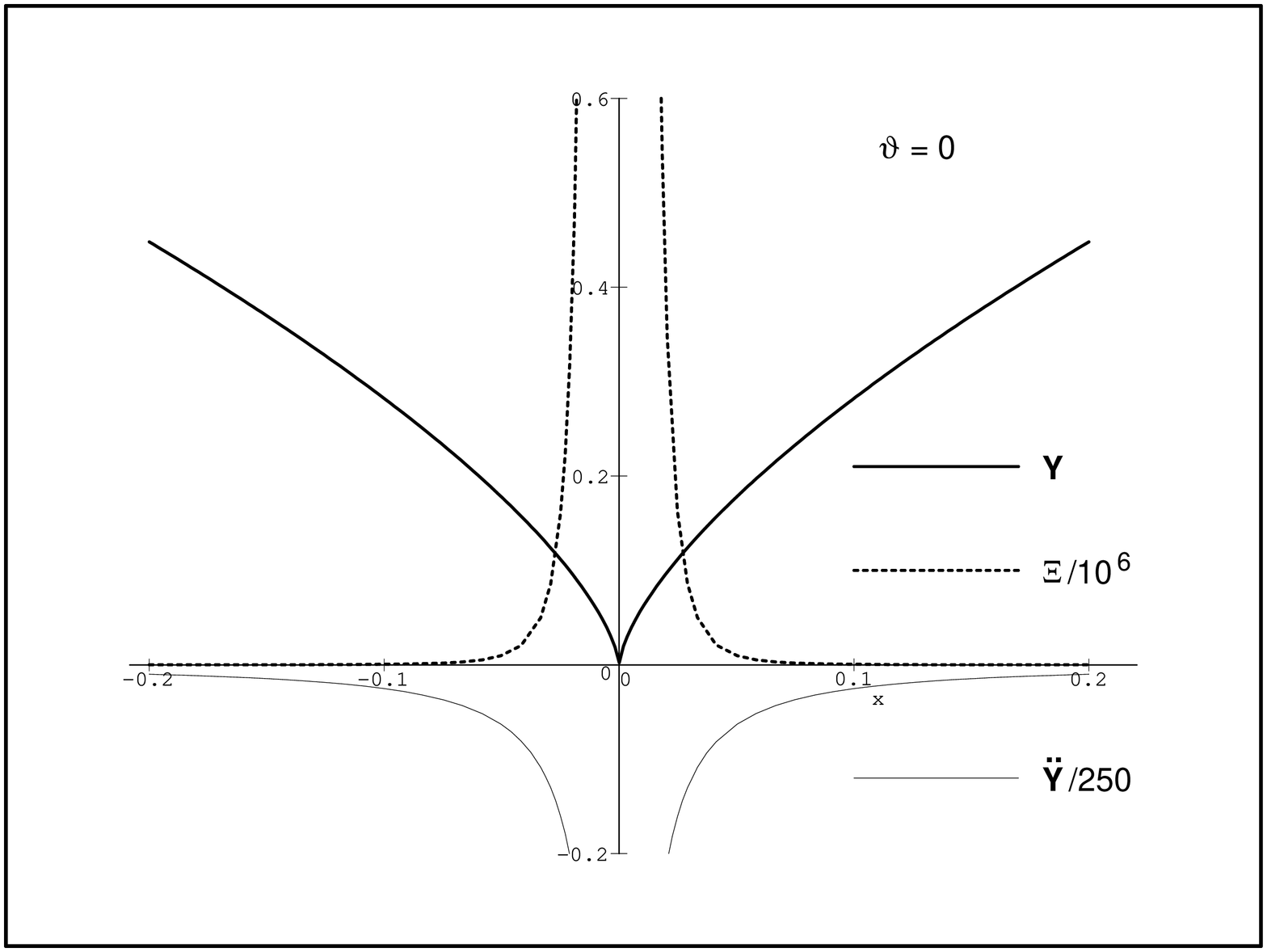,width=12cm,angle=-90}
\caption{The Early Universe for $\vartheta=0$}
\label{fig2}
The solution $\SS(x)\approx\vert{3\,x/2}\vert^{2/3}$ 
for the standard Friedmann model is 
obtained by choosing $\vartheta=0$ in \Eqn{keqn1}. The 
``big-bang'' or ``big-crunch'' singularity follows from  
$\SS\rightarrow0$ and $\Xi\rightarrow\infty$ as $x\rightarrow0$.
$\ddot{\SS}(x)$ is negative definite throughout  
spacetime.
\end{figure}

\begin{figure}
\epsfig{file=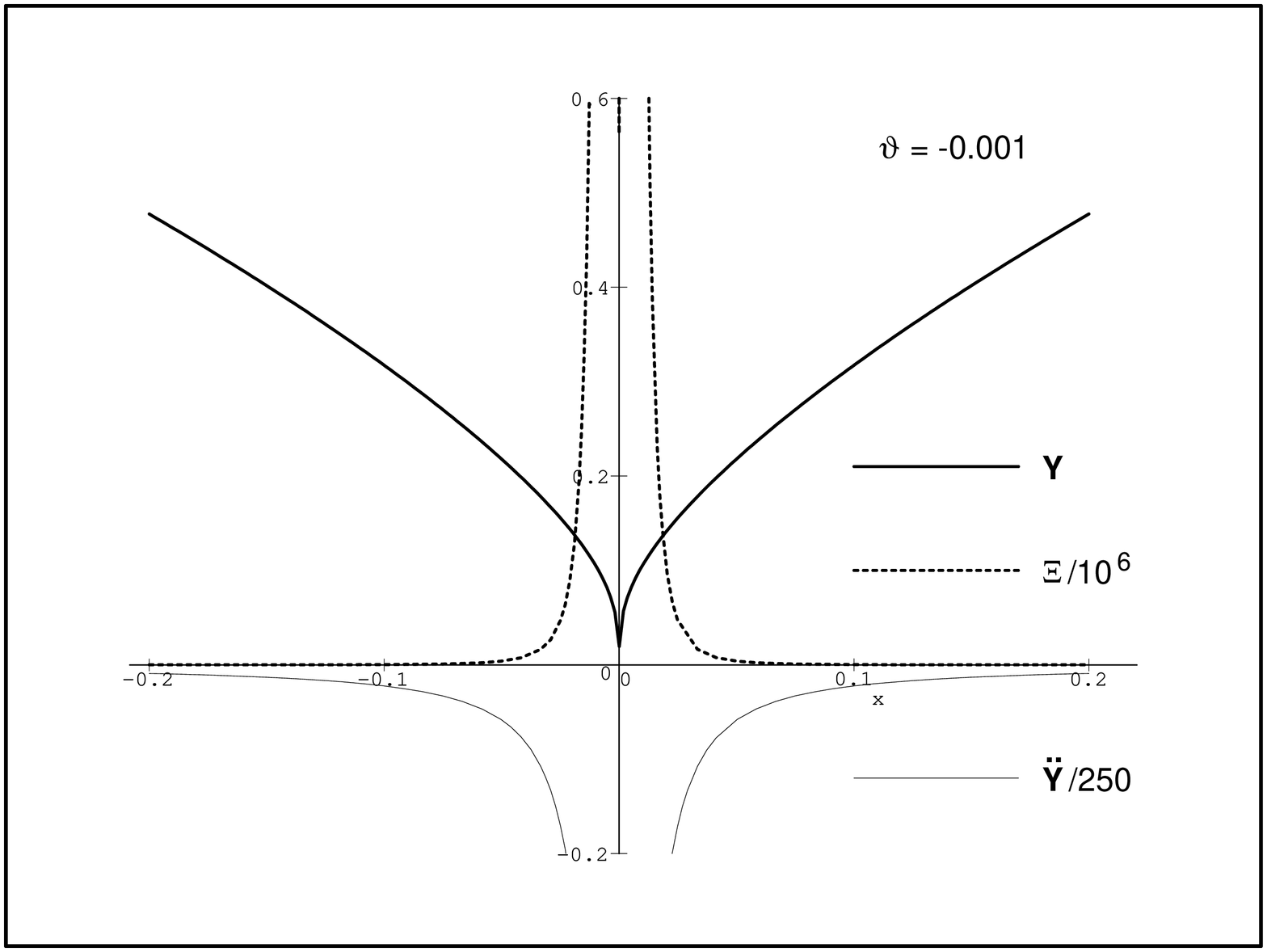,width=12cm,angle=-90}
\caption{The Early Universe for $\vartheta<0$}
\label{fig3}
With a choice of negative $\vartheta$
\Eqn{keqn1} yields the asymptotic solution
$\SS(x)\approx\vert\vartheta\vert^{1/6}\,
\vert3\,x\vert^{1/3}$ which approaches zero as 
$x\rightarrow0$  more rapidly than
$\SS(x)$ described by the Friedmann solutions, thus leading to a more 
severe singular behaviour of the spacetime curvature.
\end{figure}

In the following it is convenient to
introduce the constant $\Theta$ by
\beqn
\Theta={\sigma\over 16\,\pi\,G}
\left({\hbar\,q\,N_q\over \,m_\alpha}\right)^2
.\eeqn{Theta}
Note that the sign of $\Theta$ is determined by the coupling 
polarity $\sigma$.
In terms of the dimensionless quantities defined by
$$
x={3\,c^3\,t\over8\,\pi\,G\,M}
$$
$$
\vartheta={27\over512}\,{c^6\,\Theta\over\pi^3\,G^3\,M^4}
$$
and
$$
\SS(x)={3\,c^2\,S(t(x))\over8\,\pi\,G\,M}
$$
\Eqn{keyeqn} may be expressed as
\beqn
\left({\partial \SS\over \partial x}\right)^2
-{1\over \SS}+{\vartheta\over \SS^4}
+k=0
\eeqn{keqn1}
The solutions of this equation for various choices of $\vartheta$
(and $k=0$) are
illustrated in Figures  \ref{fig1}, \ref{fig2}, \ref{fig3} 
around $x=0$ where the solution with
small $\vartheta$ is
insensitive to the choice of $k$.

For $\sigma=-1$ where $\Theta < 0$ one finds solutions for  $\SS$
exhibiting
zeroes where the Levi-Civita geometry is singular as in the standard
model of cosmology.
For the case $\sigma=+1$ where $\Theta > 0$
any positive valued solution to \Eqn{keqn1} has a non-zero
 minimum $\SS_{\rm min}$
given by ${\partial\SS\over\partial x}|_{\SS=\SS_{\rm min}}=0$.
If $0 < \vartheta \ll 1$ then
\beqn
\SS_{\rm min}\approx
\vartheta^{1/3}.
\eeqn{SSmin}
For
$\SS_{\rm min}$ non-zero the scale factor never vanishes and the
geometry is non-singular. For example
the Levi-Civita quadratic curvature invariant
\beqn
\rinv\equiv
\star^{-1}(\RFMLC{}^a{}_b \wdg\star \RFMLC{}^b{}_a)
\eeqn{defrinv}
has the value
\beqn
\rinv
=
-{243\,c^8\over512\,\pi^4\,G^4\,M^4}
\left(
{5\over 16\,\SS^6}
-{\vartheta\over \SS^9}
+{5\,\vartheta^2\over 4\,\SS^{12}}
\right)
\eeqn{frrinv}
which is clearly finite for $Y > 0$. 

\begin{figure}
\epsfig{file=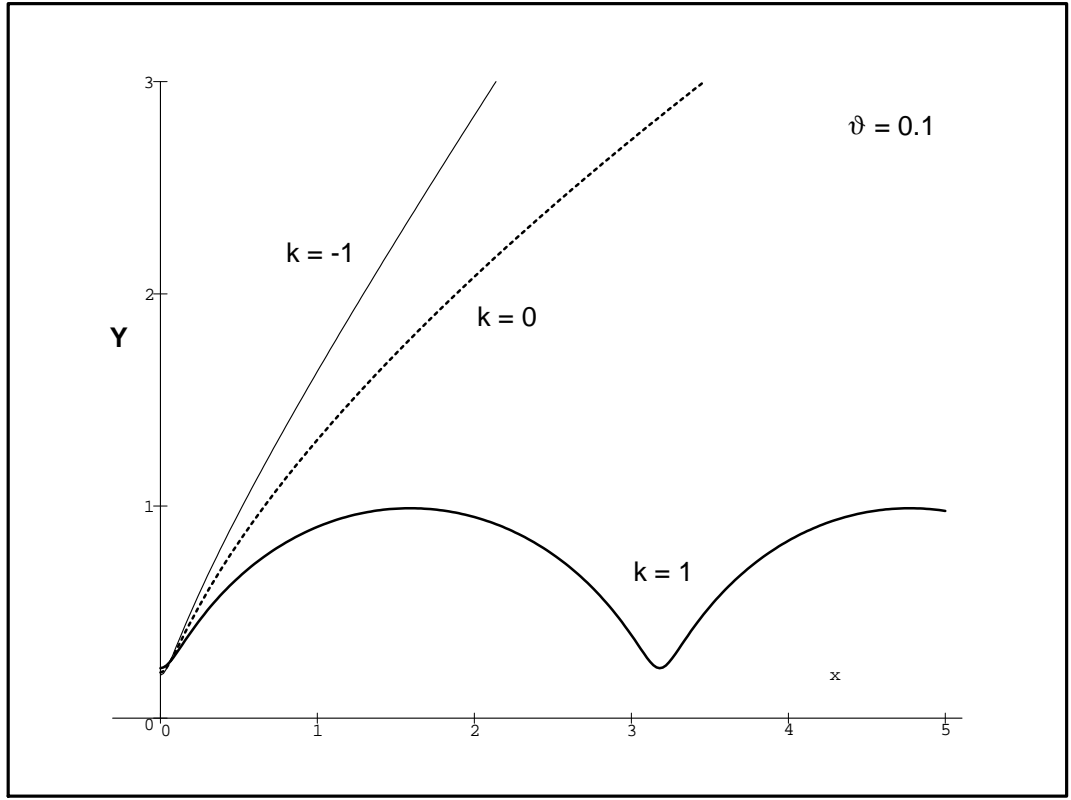,width=12cm,angle=-90}
\caption{Eternal and Oscillating Cosmologies}
\label{fig4}
The global behaviour of $\SS(x)$ for 
$\vartheta=0.1$ exhibits three types of
eternal cosmology. If the curvature parameter $k=1$
an oscillating universe is depicted by a periodic  
$\SS(x)$. If $k=0$ or $-1$ the universe contracts to a minimal scale and then
expands to infinity. 
In all three cases the Levi-Civita curvature is regular throughout
spacetime.
\end{figure}

The invariant $\rinv$ has been scaled to
\beqn
\Xi=
-{512\,\pi^4\,G^4\,M^4\over243\,c^8}\,\rinv
\eeqn{defXi}
for display in Figures \ref{fig1}, \ref{fig2}, \ref{fig3}  
in the vicinity of  $x=0$. 
We also show that in the domain 
$
\SS(x) < 4^{1/3}\,\vartheta^{1/3}
$,
$\SS(x)$ has an acceleration phase
(${\partial^2 \SS\over \partial x^2} > 0\, ,q_d < 0$ 
obtained  by differentiating \Eqn{keqn1}).
In Figure \ref{fig4} the  nature of the scale factor is exhibited over a larger
region of $x$. The eternal nature of the cosmology is apparent for
$\vartheta$ positive although depending on the size of $\Smin$
the model can readily accommodate hot dense phases where quantum
conditions may be relevant.

\vfill\eject

\Section{Dynamical Modifications to Inflationary Friedmann Cosmologies}


In the above we have shown that the presence of the Proca field in the
Einstein equations with $\sigma=1$ can modify the singular nature of the
geometry associated with the Friedmann cosmologies. It is therefore of
interest to examine how this field modifies these cosmologies when they are
preceded by an inflationary expansion era.


We recall the motivation for an inflationary phase starting at some $t=t_i$
after the standard singularity at $t=0$. If $t_i$ is chosen to be the 
Planck time $t_{\rm pl}$ then from
 \Eqn{defDelta} a
  Friedmann radiation era predicts
\beqn
\tau\equiv {\Delta_0\over\Delta_i}\approx {t_0\over t_{\rm pl}}\approx 10^{60}
\eeqn{10e60}
where $\Delta_0=\Delta(t_0)$ for $t_0\approx3.5\times10^{17}\;{\rm
sec}$ and  $t_{\rm pl}\approx 5.3\times10^{-44}\;{\rm
sec}$. 
It is accepted wisdom that
unnatural  constraints  are required to maintain this ratio over such
a long period.
The standard inflationary model
\cite{guth,guth_pi,Albrecht_Steinhardt,Hawking_Moss,linde1,linde2,llkc}
 attempts to bring the ratio $\tau$ to order 1
by postulating a pre-Friedmann era from $t_i$ to $t_r$ in which
the scale factor grows exponentially. 
Thus if the inflationary  epoch ends at $t=t_r$ with an  \efolding\ 
$\epsilon$ at an energy scale
$\sim 10^{15}$ GeV:  
\cite{frieman}
\beqn
\tau= 
{\Delta_0\over\Delta_r}\,
{\Delta_r\over\Delta_i}\approx
e^{2(60-\epsilon)}
\eeqn{e60}
By choosing $\epsilon>  {60}$ the fine tuning puzzle is considered to be
solved.

We first note that the parameters of the purely inflationary models enable one
to gain some initial estimates
of $\Theta$. 
Standard inflationary models
postulate that during the inflationary phase ($t>t_i$)
the Universe
is dominated by the constant mass density 
$\rho=\rho_v\approx 1\times10^{77}$ Kg/m$^3$ 
which drives a rapid expansion of 
$S$ 
until 
$S=S_r\approx10$ m  \cite{fowler}.
Thus for the pure inflation phase we
prescribe the equation of state
\beqn
P=-c^2\,\rho
\eeqn{infeqstat}
so by differentiating \Eqn{eineq1} and using 
\Eqn{eineq2}  the Proca modified
Einstein equations yields
\beqn
\rho={\Lambda\over 8\,\pi\,G}=\rho_v
\eeqn{effcosmosconst}
and
\beqn
\left({\partial S\over \partial t}\right)^2
-{\Lambda\over3}\,S^2
+{8\,\pi\,G\over3}\,{\Theta\over S^4}
+c^2\,k=0
\eeqn{infkeyeqn}
where the constant $\Lambda$ is the ``effective cosmological constant''.
(Standard inflation is described by  \Eqn{infkeyeqn} with $\Theta=0$.)
With ${\partial S\over \partial t}=0$ at $S=\Smin$ then
\beqn
\Theta={1\over8\pi\,G}
\left({\Lambda\,\Smin^6-3\,k\,c^2\,\Smin^4}\right).
\eeqn{LamSmin}
Thus with
$\log{\left({S_r\over \Smin}\right)}>60$
\beqn
\Smin\equiv S(t_i)<10^{-25}\,{\rm m}
\eeqn{Smin_val}
and \Eqn{LamSmin} implies that
\beqn
\Theta<5\times10^{-73} \;{\rm Kg}\,{\rm m}^3
\eeqn{Theta_value}
for $k=0,1,-1$.
If the current Friedmann era of the Universe with 
$t=t_0\approx3.5\times10^{17}\;{\rm sec}$ 
is dominated by dark matter with
negligible pressure ($\rho_q(t_0) \ll P_q(t_0)/c^2$), 
then the current scale factor may be
estimated \cite{fowler} to be 
$S(t_0)\approx3\times10^{10}\;{\rm light\;years}=3\times10^{26}\,{\rm m}$. 
Furthermore,
\beqn
M_q\approx M \approx\rho_{\rm c}(t_0)\,S(t_0)^3=2\times10^{54}\,{\rm Kg}
\eeqn{M_q}
where the critical density has been chosen to be 
$6\times10^{-26} {\rm Kg/m^3}$. 
It then follows from \Eqn{M_q}, \Eqn{muvalue}
and
\beqn
N_q={M_q\over\mu_q}\approx 1.3\times10^{77} \;{\rm m^{-3}}
\eeqn{Nq_val}
that
\beqn
{q^2\over m_\alpha{}^2}
=16\,\pi\,G\left({1\over\hbar\,N_q}\right)^2\,\Theta
<10^{-168} \;{\rm m^2/Kg^2}.
\eeqn{q_m_val}
If $m_\alpha$ is less than 
the Planck mass 
$m_{\rm pl}=\sqrt{\hbar\,c\over G}= 
2.18\times10^{-8}$~Kg then  
$\vert q \vert < 2\times10^{-92}$~m which is much smaller than the
Planck length $\ell_{\rm pl}=\sqrt{\hbar\,G\over c^3}=1.6 \times
10^{-35} $ m and somewhat unnatural.
Furthermore the dimensionless coupling ${\hat{q}^2\over\hbar\,c}$
associated with  the effective 
coupling
\beqn
\hat{q}\equiv{c^2\,q\over\pi\sqrt{32\,G}}
\eeqn{hatq}
is many orders of magnitude smaller than the fine structure constant. 
The origin of these values is the high $e$-folding $\epsilon$ needed in
the standard inflationary model to accommodate the flatness problem
associated with the standard  Friedmann curvature singularity. However
in the model with Proca fields  there need be no such singularity
and one may adjust the size of the minimum scale factor to accommodate
the flatness problem with a smaller $\epsilon$. We estimate the
necessary parameters by requiring that
 $q\approx \ell_{\rm pl} $ 
and $m_\alpha\approx m_{\rm pl}$. In this case 
the dimensionless coupling strength between two Proca charges is
\beqn
{\hat{q}^2\over\hbar\,c}
={1\over32\pi^2}\approx{1\over315}
\eeqn{procastreng}
which is of the order of fine structure constant.

For simplicity we  arrange that at the end of the inflationary
epoch the Universe proceeds to the current time in a Friedmann era
dominated by dark matter modelled as a Proca charged fluid with
negligible pressure.
From~\Eqn{LamSmin} the minimal value of the scale factor 
$\Smin=S(0)$
may be determined by $\Theta$ and $\Lambda$
\beqn
\Smin=
\sqrt{2}\,\left({\pi\,G\,\Theta\over\Lambda}\right)^{1/6}
\eeqn{smin}
neglecting $k$.
If the inflationary era ends at time $t_r$ with a scale factor 
$S_r=S(t_r)$
and  
\beqn
\rho_{\rm inflation}(t_r)=\rho_{\rm Friedmann}(t_r)
\eeqn{contrho}
one has
\beqn
\Lambda={8\,\pi\,G\,M_q\over S_r{}^3}.
\eeqn{Lamb}
With the value of $N_q$ from \Eqn{Nq_val} and  $q\approx \ell_{\rm pl} $ 
, $m_\alpha\approx m_{\rm pl}$, \Eqn{Theta} gives
\beqn
\Theta=3\times10^{24}\;{\rm Kg\,m^3}.
\eeqn{newtheta}
It follows from \Eqn{defDelta},  \Eqn{rhoq}, 
\Eqn{M_q}  and \Eqn{contrho}
that
\beqn
\tau=
{\Delta_0\over\Delta_r}\,{\Delta_r\over\Delta_i} = {S_0\over
S_r}\left(\Smin\over S_r\right)^2 \eeqn{tau} Eliminating $\Smin$ using
\Eqn{smin} and assuming the desired value $\tau< 1$ we may solve
\Eqn{tau} for $S_r$: 
\beqn 
S_r>2\times 10^8\;{\rm m}
\eeqn{srval} 
Inserting this value in \Eqn{smin} yields the value of
the minimal scale factor 
\beqn 
\Smin> 10\;{\rm cm}
\eeqn{numer_srval} 
This corresponds to 
$\epsilon=\log{({S_r/\Smin})}> 20$ which is a considerably smaller bound 
than the traditional
inflationary expansion based on Einsteinian gravitation alone.

\Section{Conclusion}

Motivated by the structure of a class of actions that involve (in
addition to the generalised Einstein-Hilbert action) terms including
the torsion and metric gradient of a general connection on the bundle
of linear frames over spacetime, the consequences of Einstein-Proca
gravitation coupled to matter have been examined. This theory is
written entirely in terms of the traditional torsion free, metric
compatible connection where all the effects of torsion and
non-metricity reside in a single vector field satisfying the Proca
equation.  In such a theory the weak field limit admits both massless
tensor gravitational quanta (traditional gravitons) and massive vector
gravitational quanta.  The mass of the Proca field is determined by
the coupling constants in the parent non-Riemannian action.  The
interaction mediated by the new Proca component of gravitation is
expected to modify the traditional gravitational interaction on small
scales.  In order to confront this expected modification with
observation we have constructed an Einstein-Proca-Fluid model in which
the matter is regarded as a perfect thermodynamic fluid. We have
suggested that in addition to ordinary matter that couples
gravitationally through its mass the conjectured dark matter in the
Universe may couple gravitationally through both its mass and a new
kind of gravitational charge. The latter coupling is analogous to the
coupling of electric charge to the photon where the analogue of the
Maxwell field is the Proca field strength (the curl of the Proca
field).
If one assumes that 
the amount of dark matter dominates over the ordinary matter in the
later phase of evolution of the Universe, that
the Proca field mass is of the order of the Planck
mass and the appropriate coupling to the dark matter is of the same
order as the fine structure constant then one finds that such  hypotheses
are consistent with both the inflationary scenario of modern cosmology
as well as the observed galactic rotation curves according to
Newtonian dynamics.  The latter follows by assuming that stars,
composed of ordinary (as opposed to dark matter), interact via
Newtonian forces to an all pervading background of massive
gravitationally charged cold dark matter in addition to ordinary matter.  
The novel gravitational
interactions are predicted to have a significant influence on
pre-inflationary cosmology.  For attractive forces between dark matter
charges of like polarity the Einstein-Proca-matter system exhibits
homogeneous isotropic eternal cosmologies that are free of
cosmological curvature singularities thus eliminating the horizon
problem associated with the standard big-bang scenario. Such solutions
do however admit dense hot pre-inflationary epochs each with a
characteristic scale factor that may be correlated with the dark
matter density in the current era of expansion.

The Einstein-Proca-Fluid model offers a simple phenomenological
description of dark matter gravitational interactions. It has its
origins in a geometrical description of gravitation and the theory
benefits from a variational formulation in which the connection is a
bona fide dynamical variable along with the metric. The simplicity of
the model is a consequence of the structure of a class of
non-Riemannian actions whose dynamical consequences imply that the new
physics resides in a component of gravitation mediated by a Proca
field. It will be of interest to confront the theory with other aspects
of astrophysics such as localised gravitational collapse, the
nature of the inflation mechanism and the  origin of dark matter.

\vfill\eject

\Section{Acknowledgement}

The authors are grateful to J Schray and D H Lyth for valuable interaction.
RWT is grateful to the Human Capital and Mobility Programme of the European
Union for partial support. CW is grateful to the Committee of
Vice-Chancellors and Principals, UK for an Overseas Research Studentship and
to Lancaster University for a School of Physics and Chemistry Studentship
and a Peel Award.




\end{document}